\documentclass[prd,showpacs]{revtex4}
\begin{document}

\newcommand{\re}{\mathop{\mathrm{Re}}}

\newcommand{\be}{\begin{equation}}
\newcommand{\ee}{\end{equation}}
\newcommand{\bea}{\begin{eqnarray}}
\newcommand{\eea}{\end{eqnarray}}

\title{Canonical superenergy tensors in general relativity: a reappraisal}

\author{Janusz Garecki}
\email{garecki@wmf.univ.szczecin.pl}
\affiliation{\it Institute of Mathematics University of Szczecin
and Cosmology Group University of Szczecin,
 Wielkopolska 15, 70-451 Szczecin, Poland}
\date{\today}

\input epsf

\begin{abstract}
Here we present our actual point of view on the canonical superenergy tensors.
\end{abstract}


\maketitle

\section{Introduction}

In the framework  of general relativity ({\bf GR}), as a consequence of the Einstein Equivalence
Principle ({\bf EEP}, the gravitational field {\it has non-tensorial strengths} $\Gamma^i_{kl}
 = \{^i_{kl}\}$ and {\it admits no energy-momentum tensor}. One
 can only attribute to this field {\it gravitational
 energy-momentum pseudotensors}. The leading object of such a kind
 is the {\it canonical gravitational energy-momentum pseodotensor}
 $_E t_i^{~k}$ proposed already in past by Einstein. This
 pseudotensor is a part of the {\it canonical energy-momentum
 complex} $_E K_i^{~k}$ in {\bf GR}.

The canonical complex $_E K_i^{~k}$ can be easily obtained by
rewiriting Einstein equations to the superpotential form
\begin{equation}
_E K_i^{~k} := \sqrt{\vert g\vert}\bigl( T_i^{~k} + _E
t_i^{~k}\bigr) = _F U_i^{~[kl]}{}_{,l}
\end{equation}
where $T^{ik} = T^{ki}$ is the symmetric energy-momentum tensor for matter, $g = det[g_{ik}]$,
 and

\begin{eqnarray}
_E t_i^{~k}& =& {c^4\over 16\pi G} \bigl\{\delta_i^k
g^{ms}\bigl(\Gamma^l_{mr}\Gamma^r_{sl} -
\Gamma^r_{ms}\Gamma^l_{rl}\bigr)\nonumber\cr
&+& g^{ms}_{~~,i}\bigl[\Gamma^k_{ms} - {1\over 2}
\bigl(\Gamma^k_{tp}g^{tp} -
\Gamma^l_{tl}g^{kt}\bigr)g_{ms}\nonumber\cr
&-& {1\over 2}\bigl(\delta^k_s \Gamma^l_{ml} +
\delta^k_m \Gamma^l_{sl}\bigr)\bigr]\bigr\};
\end{eqnarray}
\begin{equation}
_F {U_i^{~[kl]}} = {c^4\over 16\pi G}g_{ia}({\sqrt{\vert
g\vert}})^{(-1)}\bigl[\bigl(-g\bigr)\bigl(g^{ka} g^{lb} - g^{la}
g^{kb}\bigr)\bigr]_{,b}.
\end{equation}
$_E t_i^{~k}$ are components of the canonical energy-momentum
pseudotensor for gravitational field $\Gamma ^i_{kl} =
\bigl\{^i_{kl}\bigr\}$, and $_F {U_i^{~[kl]}}$ are von Freud
superpotentials.
\begin{equation}
_E K_i^{~k} = \sqrt{\vert g\vert}\bigl(T_i^{~k} + _E
t_i^{~k}\bigr)
\end{equation}
are components of the {\it Einstein canonical energy-momentu complex,
for matter and gravity}, in {\bf GR}.

Symbol $,l$ means partial derivative $\partial_l$.

In the consequence of (1) the complex $_E K_i^{~k}$satisfies local
conservation laws
\begin{equation}
{_E K_i^{~k}}_{,k}\equiv 0.
\end{equation}

In very special cases one can obtain from these local conservation
laws reasonable integral conservation laws.

Despite that one can easily introduce in {\bf GR} {\it the
canonical (and others) superenergy tensor} for gravitational
field. This was done in past in a series of our articles (See,
e.g.,\cite{Gar1} and references therein).

It appeared that the idea of the superenergy tensors is universal:
to any physical field having an energy-momentum tensor or
pseudotensor one can attribute the coresponding superenergy
tensor.
\section{The canonical superenergy tensors}

Here we give a short reminder of the general, constructive
definition of the superenergy tensor $S_a^{~b}$ applicable to
gravitational field and to any matter field. The definition uses
{\it locally Minkowskian structure} of the spacetime in {\bf GR}
and, therefore,
it fails in a spacetime with torsion, e.g., in Riemann-Cartan
spacetime.

In the normal Riemann coordinates {\bf NRC(P)} we define
(pointwiese)
\begin{equation}
S_{(a)}^{~~~(b)}(P) = S_a^{~b} :=(-) \displaystyle\lim_{\Omega\to
P}{\int\limits_{\Omega}\biggl[T_{(a)}^{~~~(b)}(y) - T_{(a)}^{
~~~(b)} (P)\biggr]d\Omega\over 1/2\int\limits_{\Omega}\sigma(P;y)
d\Omega},
\end{equation}
where
\begin{eqnarray}
T_{(a)}^{~~~(b)}(y) &:=& T_i^{~k}(y)e^i_{~(a)}(y)
e_k^{~(b)}(y),\nonumber\cr
T_{(a)}^{~~~(b)}(P)&:=& T_i^{~k}(P) e^i_{~(a)}(P)e_k^{~(b)}(P) =
T_a^{~b}(P)
\end{eqnarray}
are {\it physical or tetrad components} of the pseudotensor or
tensor field which describes an energy-momentum distribution, and $\bigl\{y^i\bigr\}$
are normal coordinates. $e^i_{~(a)}(y), e_k^{~(b)} (y)$ mean an
orthonormal tetrad $e^i_{~(a)}(P) = \delta_a^i$ and its dual $e_k^{~(a)}(P) = \delta_k^a $
paralelly propagated along geodesics through $P$ ($P$ is the origin
of the {\bf NRC(P)}).

We have
\begin{equation}
e^i_{~(a)}(y) e_i^{~(b)}(y) = \delta_a^b.
\end{equation}

For a sufficiently small 4-dimensional domain $\Omega$ which
surrounds {\bf P} we require
\begin{equation}
\int\limits_{\Omega}{y^i d\Omega} = 0, ~~\int\limits_{\Omega}{y^i
y^k d\Omega} = \delta^{ik} M,
\end{equation}
where
\begin{equation}
M = \int\limits_{\Omega}{(y^0)^2 d\Omega} =
\int\limits_{\Omega}{(y^1)^2 d\Omega} =
\int\limits_{\Omega}{(y^2)^2
d\Omega}=\int\limits_{\Omega}{(y^3)^2 d\Omega},
\end{equation}
is a common value of the moments of inertia of the domain $\Omega$
with respect to the subspaces $y^i = 0,~~(i= 0,1,2,3)$.

We can take as $\Omega$, e.g., a  sufficiently small analytic ball centered
at $P$:
\begin{equation}
(y^0)^2 + (y^1)^2 + (y^2)^2 + (y^3)^2 \leq R^2,
\end{equation}
which for an auxiliary positive-definite metric
\begin{equation}
h^{ik} := 2 v^i v^k - g^{ik},
\end{equation}
can be written in the form
\begin{equation}
h_{ik}y^i y^k \leq R^2.
\end{equation}
A fiducial observer {\bf O} is at rest at the beginning {\bf P}
of the used Riemann normal coordinates {\bf NRC(P)} and its four-
velocity is $v^i =\ast~ \delta^i_o.$ $=\ast$ means that an
equations is valid only in special coordinates.

$\sigma(P;y)$ denotes the two-point {\it world function}
introduced in past by J.L. Synge \cite{Synge}
\begin{equation}
\sigma(P;y) =\ast {1\over 2}\bigl(y^{o^2} - y^{1^2} - y^{2^2}
-y^{3^2}\bigr).
\end{equation}

The world function $\sigma(P;y)$ can be defined covariantly by the
{\it eikonal-like equation} \cite{Synge}
\begin{equation}
g^{ik} \sigma_{,i} \sigma_{,k} = 2\sigma,
~~\sigma_{,i} := \partial_i\sigma,
\end{equation}
together with
\begin{equation}
\sigma(P;P) = 0, ~~\partial_i\sigma(P;P) = 0.
\end{equation}

The ball $\Omega$ can also be given by the inequality
\begin{equation}
h^{ik}\sigma_{,i} \sigma_{,k} \leq R^2.
\end{equation}

Tetrad components and normal components are equal at {\bf P}, so,
we will write the components of any quantity attached to {\bf P}
without tetrad brackets, e.g., we will write $S_a^{~b}(P)$
instead of $S_{(a)}^{~~~(b)}(P)$ and so on.

If $T_i^{~k}(y)$ are the components of an energy-momentum tensor
of matter, then we get from (5)
\begin{equation}
_m S_a^{~b}(P;v^l) = \bigl(2{\hat v}^l {\hat v}^m - {\hat g}^{lm}\bigr) \nabla_l \nabla_m {}
{\hat T}_a^{~b} = {\hat h}^{lm}\nabla_l \nabla_m {}{\hat T}_a^{~b}.
\end{equation}
Hat over a quantity denotes its value at {\bf P}, and $\nabla$
means covariant derivative.

Tensor $_m S_a^{~b}(P;v^l)$ is {\it the canonical superenergy tensor for matter}.

For the gravitational field, substitution of the canonical
Einstein energy-momentum pseudotensor as $T_i^{~k}$ in (5) gives
\begin{equation}
_g S_a^{~b}(P;v^l) = {\hat h}^{lm} {\hat W}_a^{~b}{}_{lm},
\end{equation}
where
\begin{eqnarray}
{W_a^{~b}}{}_{lm}&=& {2\alpha\over 9}\bigl[B^b_{~alm} +
P^b_{~alm}\nonumber\cr
&-& {1\over 2}\delta^b_a R^{ijk}_{~~~m}\bigl(R_{ijkl} +
R_{ikjl}\bigr) + 2\delta_a^b{\beta}^2 E_{(l\vert g}{}E^g_{~\vert
m)}\nonumber\cr
&-& 3 {\beta}^2 E_{a(l\vert}{}E^b_{~\vert m)} + 2\beta
R^b_{~(a\vert g\vert l)}{}E^g_{~m}\bigr].
\end{eqnarray}
Here $\alpha = {c^4\over 16\pi G} = {1\over 2\beta}$, and
\begin{equation}
E_i^{~k} := T_i^{~k} - {1\over 2}\delta_i^k T
\end{equation}
is the modified energy-momentum tensor of matter \footnote{In
terms of $E_i^{~k}$ Einstein equations read $R_i^{~k} = \beta
E_i^{~k}$.}.

On the other hand
\begin{equation}
B^b_{~alm} := 2R^{bik}_{~~~(l\vert}{}R_{aik\vert m)}-{1\over
2}\delta_a^b{} R^{ijk}_{~~~l}{}R_{ijkm}
\end{equation}
are the components of the {\it Bel-Robinson tensor} ({\bf BRT}),
while
\begin{equation}
P^b_{~alm}:= 2R^{bik}_{~~~(l\vert}{}R_{aki\vert m)}-{1\over
2} \delta_a^b{}R^{jik}_{~~~l}{}R_{jkim}
\end{equation}
is the Bel-Robinson tensor with  ``transposed'' indices $(ik)$.

Tensor $_g S_a^{~b}(P;v^l)$ is the {\it canonical superenergy
tensor} for gravitational field $\bigl\{^i_{kl}\bigr\}$.

In vacuum $_g S_a^{~b}(P;v^l)$ takes the simpler form
\begin{equation}
_g S_a^{~b}(P;v^l) = {8\alpha\over 9} {\hat h}^{lm}\bigl({\hat
C}^{bik}_{~~~(l\vert}{}{\hat C}_{aik\vert m)} -{1\over
2}\delta_a^b {\hat C}^{i(kp)}_{~~~~~(l\vert}{}{\hat C}_{ikp\vert
m)}\bigr).
\end{equation}
Here $C^a_{~blm}$ denote components of the {\it Weyl tensor}.

Some remarks are in order:
\begin{enumerate}
\item in vacuum the quadratic form $_g S_a^{~b} v^a v_b$, where $v^a v_a = 1$
{\it is positive-definite}. This form gives the gravitational {\it superenergy
density} $\epsilon_g$ for a fiducial observer {\bf O}.
\item In general, the canonical superenergy tensors are uniquely
determined only along the world line of the observer {\bf O}. But
in special cases, e.g., in Schwarzschild spacetime or in Friedman
universes, when there exists a physically and geometrically
distinguished  four-velocity field $v^i(x)$, one can introduce in
an unique way  the unambiguous fields $_g S_i^{~k}(x;v^l)$ and $_m
S_i^{~k}(x;v^l)$.
\item It can be shown that the superenergy densities $\epsilon_g,
~~\epsilon_m$, which have dimension ${Joul\over metre^{5}}$,
exactly corespond to the Appel's {\it energy of acceleration} ${1\over 2}{\vec a}{\vec
a}$.

The Appel's energy of acceleration plays fundamental role in Appel's approach
to classical mechanics \cite {Appel}.
\item
We have proposed in our previous papers to use the tensor $_g S_i^{~k}(P;v^l)$
as a substitute of the non-existing gravitational energy-momentum
tensor.
\item In past we have used the canonical superenergy  tensors $_g S_i^{~k}$
and $_m S_i^{~k}$ to local (and also to global) analysis of some
well-known solutions to the Einstein equations like Schwarzschild,
Kerr, Friedman, G\"odel, Kasner, Bianchi I, de Sitter and anti-de
Sitter solutions. The obtained results were very interesting
(See,\cite{Gar1}), e.g., in G\"odel universes the sign of the
superenergy density $\epsilon_s := \epsilon_g + \epsilon_m$
depends on causality $(\epsilon_s <0)$ and non-causality $(\epsilon_s
>0)$, and, in Schwarzschild spacetime the integral exterior
superenergy $S$ is connected with Hawking temperature $T$ of the
Schwarzschild black hole: $S = {8\pi k c^3\over 9\hbar G} T$.
We have also studied the transformational rules for the canonical
superenergy tensors under conformal rescaling of the metric $g_{ik}(x)$
\cite {Gar1,Gar2}.
\item The idea of the superenergy tensors can be extended on
angular momentum (See, \cite{Gar1}). The obtained angular supermomentum
tensors {\it do not depend} on a radius vector and, in gravitational case, they depend
only on ``spinorial part'' of the suitable gravitational angular
momentum pseudotensor.
\item As a result of an averaging the tensors $_g S_a^{~b}(P;v^l)$
and $_m S_a^{~b}(P;v^l)$, in general, do not satisfy any local
conservation laws. Only in a symmetric spacetime or in a spacetime
which has constant curvature one can get
\begin{equation}
\bigl[_g S_a^{~b}(P;v^l)\bigr]_{,b} = 0.
\end{equation}
\item There exists exchange of the canonical superenergy  between
gravity and matter in the following sense. Let us consider the
consequence of the equations (4)
\begin{equation}
\bigl(\Delta^{(4)}_E K_i^{~k}\bigr)_{,k} =\bigl[ \bigl( \Delta^{(4)}
(\sqrt{\vert g\vert} _E t_i^{~k}\bigr)+ \Delta^{(4)}\bigl(\sqrt{\vert
g\vert} T_i^{~k}\bigr)\bigr]_{,k} = 0,
\end{equation}
where $\Delta^{(4)} := (\partial_0)^2 + (\partial_1)^2 + (\partial_2)^2
+(\partial_3)^2$.

The exchanged quantities (with total balance equal to zero)
\begin{equation}
\Delta^{(4)}\bigl(\sqrt{\vert g\vert}_e t_i^{~k}\bigr),
~~\Delta^{(4)}\bigl(\sqrt{\vert g\vert} T_i^{~k}\bigr)
\end{equation}
have dimensions of the canonical superenergy and, when taken at the
beginning {\bf P} of the {\bf NRC(P)} and written covariantly, then they coincide with
the canonical superenergy tensors $ _g S_i^{~k}(P;v^l), ~~_m S_i^{~k}(P;v^l)$
respectively.
\end{enumerate}

Changing the constructive definition (5) to the form
\begin{equation}
<T_a^{~b}(P)>:= \displaystyle\lim_{\varepsilon\to
0}{\int\limits_{\Omega}{\bigl[T_{(a)}^{~~~(b)}(y) -
T_{(a)}^{~~~(b)}(P)\bigr]d\Omega}\over\varepsilon^2/2\int\limits_{\Omega}d\Omega},
\end{equation}

where $\varepsilon:= {R\over L}>0$ (equivalently $R = \varepsilon L$) is a real parameter
and $L$ is a dimensional constant:$[L] = m$, one obtains    {\it the averaged
relative energy-momentum  tensors}. Namely, from (25) one obtains:

for matter
\begin{equation}
<_m T_a^{~b}(P;v^l)> = _m S_a^{~b}(P;v^l) {L^2\over 6},
\end{equation}

and for gravity
\begin{equation}
<_g t_a^{~b}(P;v^l)> = _g S_a^{~b}(P;v^l){L^2\over 6}.
\end{equation}

The components of the averaged relative energy-momentum tensors have correct
dimensions, i.e., they have the same dimensions as the components
of an energy-momentum tensor but they depend on a dimensional
parameter $L$. So, introducing of the tensors of such a kind leads
us to serious problem, how to choose the dimensional parameter $L$?

It is seen from (26) and (27) that the averaged tensors $<_m T_a^{~b}(P;v^l)>$
and $<_g t_a^{~b}(P;v^l)>$, for matter and gravitation, can be interpreted as {\it fluxes}
of the appropriate canonical superenergy.

In the paper \cite{Gar1} we have proposed an universal choose
  of the parameter $L$. Namely, we have proposed $L = 100 L_P =\approx 10^{-33}
  m$. Here $L_P  := \sqrt{{\hbar G\over c^3}}=\approx 10^{-35}$ m
  is the {\it Planck length}.

Such choice of $L$ gives the averaged relative
  energy-momentum tensors which components are neglegible in
  comparison with components of an energy-momentum tensor for
  matter. In consequence, with such choice of the parameter $L$,
  these tensors play no role in evolution of the material objects
  and in evolution of the Universe.

  On the other hand the choices:
  \begin{enumerate}
  \item For a closed system of the mass $M$
  \begin{equation}
  L ={2GM\over c^2};
  \end{equation}
  \item For a gravitational wave of the length $\lambda$
  \begin{equation}
  L = \lambda;
  \end{equation}
  \item In cosmology
  \begin{equation}
  L = {2GM_U\over c^2} = {c\over H_o} = c t_o.
  \end{equation}
  \end{enumerate}

  lead us th the averaged relative energy densities of the same
  order as ordinary energy density of matter $\epsilon = T_{ik} v^i v^k$
  for an observer which four-velocity is $v^i$.
  Here $M_U, ~~H_o, ~~t_o$ mean mass of the observed part of the
  Universe, actual value of the Hubble constant and the approximated age of the
  Universe respectively.

So, in this case we have problem how to utilize the averaged relative energy-momentum
tensor for matter  $<_m T_a^{~b}(P;v^l)>$ because we already have the tensor $_m T_a^{~b}(P)$.

Of course, there exist other possibilities of choosing of the
length parameter $L$.

In consequence, now we think that the introducing of the one-parameter family of the averaged
relative energy-momentum tensors is not a good idea and that the ordinary
canonical superenergy tensors are better and more fundamental construction. The latter
tensors are unambiguous and they do not ``collide'' with any
energy-momentum tensor.

Recently we have observed the strong correlation between the sign
of the total superenergy density $\epsilon_s = \epsilon_g + \epsilon_m$
and stability of the solutions to the Einstein equations. Namely ,
we have noticed that the total superenergy density $\epsilon_s$ is
positive-definite or null for stable solution and negative-definite
for unstable solutions. Thus we think that the
following Conjecture is valid.
\begin{center}
{\bf Conjecture}
\end{center}

Sign of the total superenergy density $\epsilon_s$ determines
stability or instability of a solution to the Einstein equations:
if $\epsilon_s \geq 0$, then the solution is stable; when
$\epsilon_s<0$, then the solution is unstable.

We have not proved this Conjecture yet. Up to now we are only able to give examples
which confirm it.\footnote{We don't know any counterexample.}

\begin{center}
The examples
\end{center}

\begin{enumerate}
\item Exterior Schwarzschild:  $\epsilon_s>0$ --------- stable;
\item Kerr solution:   $\epsilon_s>0$ --------- stable;
\item Minkowski spacetime:  $\epsilon_s = 0$ ---------- stable;
\item Friedman universes:   $\epsilon_s >0$ ----------- stable;
\item Kasner  universe:   $\epsilon_s >0$ ---------- stable;
\item Bianchi I spacetime:   $\epsilon_s >0$ ---------- stable;
\item Anti-de Sitter spacetime:   $\epsilon_s <0$ ----------
unstable;
\item De Sitter spacetime:   $\epsilon_s < 0$ ---------- unstable.
\end{enumerate}

Instability of the de Sitter and anti-de Sitter spacetimes was
proved recently \cite{Garf}.

\section{Conclusion}

On the {\it superenergy level} or on the {\it averaged relative
energy-momentum} level we have no problem with suitable tensor for
gravity.

In our opinion, the canonical superenergy tensors seem more
fundamental than the corresponding averaged relative energy-momentum tensors, e.g.,
 they are independent of an dimensional factor $L$.

The canonical superenergy tensors are very useful to local
analysis of the solutions to the Einstein equations; especially to
analyse of their singularities.

Probably, these tensors give us also the very simple and powerful
method to study stability of the solutions to the Einstein
equations.

\acknowledgments

This paper was mainly supported by Polish Ministry of Science
and Higher Education  Grant No 505-4000-25-0976 (years 2011-1013).
Author also would like to thank Professor Jiri Bi\v cak for
possibility to deliver talk during the Conference ``One hundred years after
Einstein in Prague''.


\begin{thebibliography}
{99}
\bibitem{Gar1} J. Garecki, {\it Rep. Math. Phys.}, {\bf 33} (1993)
57; {\bf 40} (1997) 485; {\bf 44} (1999) 95; {\it Int. J. of
Theor. Phys.}, {\bf 34} (1995) 2259; {\it J. Math. Phys.}, {\bf 40}
(1999) 4035; {\it Ann. der Phys. (Leipzig)}, {\bf 11} (2002) 441;
{\it Ann. der Phys. (Berlin)}, {\bf 19} (2010) 263; {\it Class.
Quantum Grav.}, {\bf 19} (2002) 1; {\bf 22} (2005) 4051; {\it
Found.of Physics}, {\bf 37} (2007) 341; {\it Phys.Letters}, {\bf B
686} (2010) 6.
\bibitem{Synge} J.L. Synge, ``Relativity: the General Theory'',
North-Holland, Amsterdam 1960.
\bibitem{Appel} P. Appel, {\it Journal fuer die reine und
angewandte Mathematik}, {\bf 121} (1900) 310; {\bf 122} (1900)
205; G. Bia\l kowski, ``Classical Mechanics'', PWN, Warsaw 1975
(in Polish).
\bibitem{Gar2} J. Garecki et al., {\it Ann. der Physik (Berlin)}
{\bf 18} (2009) 13.
\bibitem{Garf} D. Garfinkle, ``AdS instability'', {\it Matter of
Gravity}, {\bf 39}, Winter 2012,p.7.; V. Emelyanov et al., ``De
Sitter spacetime instability from a nonstandard vector field'',
{\it Phys. Rev.} {\bf D 86} (2012) 027302.
\end{thebibliography}
\end{document}